\documentclass[twocolumn,nofootinbib]{revtex4-2}
\usepackage{graphicx} 
\usepackage[utf8]{inputenc}
\usepackage{amsmath}
\usepackage{amssymb}
\usepackage{braket}
\usepackage{graphicx}
\usepackage{float}
\usepackage{tikz}
\usepackage{xcolor}
\usepackage{amsmath}
\usepackage{pifont}
\usepackage{amsthm}
\usepackage{cancel}
\usepackage{tcolorbox}
\usepackage{paralist}
\usepackage{hyperref}
\usepackage[capitalize]{cleveref}
\usepackage{bbold}
\usepackage{enumitem}

\setlist[itemize]{noitemsep}
\setlist[enumerate]{noitemsep}

\newcommand{\ba}{\begin{eqnarray}}
\newcommand{\ea}{\end{eqnarray}}

\begin{document}

\title{Superluminal Quantum Reference Frames}

\author{Amrapali Sen}
\email{amrapalisen.phy@gmail.com}
\thanks{Corresponding author}
\affiliation{International Centre for Theory of Quantum Technologies, University of Gda\'nsk, Gda\'nsk, Poland}
\author{Matthias Salzger}
\email{matthias.salzger@outlook.com}
\affiliation{International Centre for Theory of Quantum Technologies, University of Gda\'nsk, Gda\'nsk, Poland}
\author{Łukasz Rudnicki}
\email{lukasz.rudnicki@ug.edu.pl}
\affiliation{International Centre for Theory of Quantum Technologies, University of Gda\'nsk, Gda\'nsk, Poland}
\newcommand{\ms}[1]{\textcolor{black!60!green}{ [#1]}}
\newcommand{\as}[1]{\textcolor{blue}{ [#1]}}
\newcommand{\lr}[1]{\textcolor{orange}{ [#1]}}

\begin{abstract}
   Particles cannot travel faster than the speed of light, nor can information. Nonetheless, this assumption has been frequently questioned over the years. This includes radical proposals like tachyons that can be observed in particle accelerators to more subtle ideas like studying superluminal causation to find classical explanations for quantum correlations. Most recently, it has been argued  [New J. Phys. \textbf{22}, 033038 (2020)] that in a world with superluminal observers local determinism is impossible, linking the two pillars of physics—quantum theory and relativity—suggesting that the latter serves as the foundation for the former. Motivated by this approach, in this work, we extend the framework of quantum reference frames to incorporate superluminal Lorentz transformations. We apply this conceptual result to examine an apparent paradox where particles acquire negative energies after undergoing a superluminal Lorentz boost and propose a resolution within our framework. A consequence of this resolution is that different observers no longer agree whether a particle is incoming or outgoing or in a superposition of the two. We further find a similar apparent paradox for the entropy and/or temperature under superluminal Lorentz boosts and resolve it in an analogous fashion. Finally, we discuss Bell experiments under superluminal quantum reference frame transformations, showing that the involved probabilities still remain conserved. By applying the toolbox of quantum reference frames, we start charting an unexplored territory of superpositions between sub-and-superluminal quantum observers which gives rise to new conceptual questions. 
\end{abstract}

\maketitle

\section{Introduction}
 
Physics is rooted in the quest for knowledge about the natural world, striving to uncover the
basic laws and principles that underlie all physical processes. The pursuit of understanding the fundamental laws of the universe has led to the development of two distinct and seemingly irreconcilable pillars: quantum theory and relativity. These theories describe the behavior of the universe at vastly different scales and have profoundly transformed our understanding
of physics. Over the past century, significant efforts have been made to develop, refine, and reconcile these two theories.

One of the key insights in this pursuit is understanding how fundamental symmetries manifest in quantum and relativistic settings. The Lorentz transformation, a cornerstone of special relativity, describes how space and time coordinates change between different inertial observers. In classical physics, it ensures the consistency of Maxwell’s equations and the constancy of the speed of light. However, in a fully quantum framework, reference frames themselves, if not considered as formal constructions, but rather, treated as objects associated with physical observers, shall also be treated as quantum systems. A particular framework with that feature is given by, so-called, quantum reference frames  \cite{giacomini2019quantum}. An immediate consequence of this framework is that notions like superposition and entanglement are defined only relative to the chosen reference frame, in the spirit of the relational description of physics \cite{ballesteros2021group, cepollaro2024sum, giacomini2019quantum, giacomini2019relativistic}. Thus, what looks like a superposition of physical systems from a particular choice of reference frame will look like entanglement when viewed from a different choice of reference frame. 

On the other hand, a mathematical derivation of the Lorentz transformations assuming just the principle of relativity and linearity \cite{parker1969faster,dragan2020quantum} yields two branches of transformations. One branch consists of the usual subluminal transformations for velocities less than the maximal speed being a free parameter of the theory (further associated with the speed of light), while the other one corresponds to superluminal transformations for velocities bigger than the speed of light. Additional physical constraints are required to rule out the superluminal branch.

A standard argument for the impossibility of superluminal particles, also known as tachyons, and superluminal observers says that they would allow for backwards-in-time signalling, thus causing causality paradoxes \cite{einstein1907uber,tolman1917velocities}.
Nevertheless, over the years, this assumption has been frequently questioned, and the idea of breaking this speed limit has popped up from time to time, both from a purely theoretical point of view as well as in an attempt to explain various phenomena. The topic has been treated on a theoretical level in \cite{bilaniuk1962meta,feinberg1967possibility,recami1986classical,rembielinski2012meta,grushka2013tachyon} to name just a few. These research programs are interesting from a number of perspectives. If one takes the position that superluminal phenomena actually exist in some manner and have explanatory power, this is obvious. But even if one’s point of view is opposed to this idea, then these are still interesting toy theories. Understanding why exactly they work or fail can potentially yield new insights into how different features of a theory constrain each other. In a similar vein, in order to confirm our current theories we need to check that what they predict to be impossible is actually so. 

These arguments often conflate between the existence of superluminal signalling, superluminal causation and the existence of superluminal Lorentz transformation/superluminal observers. The first option is very drastic, and, as such, generally considered to be impossible, while the second one is well under study \cite{vilasini2022impossibility,horodecki2019relativistic,durr2009bohmian}.

The third possibility has recently been proposed by Dragan and Ekert  \cite{dragan2020quantum} who argue that the causality issues of superluminal observers are only an apparent paradox, which vanishes if one drops the assumption of local determinism. The idea is that with large enough uncertainties, observers cannot say whether they actually observed superluminal signalling. Thus, they argue, the inherent randomness that we are familiar with from quantum theory could not just be reconciled with the theory of relativity, but the former can be understood as a consequence of the latter.  

If we take this idea seriously and admit that superluminal Lorentz transformations might be related to aspects of quantum mechanics, we shall also bring these transformations to the quantum realm. Therefore, the goal of this work is to extend the framework of quantum reference frames as a tool such that one can reason about superluminal Lorentz transformations in connection with quantum phenomena.\footnote{We advocate that reference frames can be either treated as classical systems or as quantum systems, depending on the context.} 

The structure of the present work can be summarized as follows. First, in the following two subsections we briefly review the key aspects of superluminal Lorentz transformations which back up the connection between quantum theory and superluminal observers  \cite{dragan2020quantum}, and briefly review the framework of quantum reference frames (QRF) \cite{giacomini2019relativistic}. Second, following \cite{lake2024towards} we discuss the possible joint group structure of Subluminal Lorentz Transformations (SbLT) taken together with Superluminal Lorentz Transformations (SpLT), and as a major result construct the \textit{Superluminal Quantum Reference Frame Transformations}. In light of the above, this paper is more conceptual, rather than computational. Third, we discuss a number of still conceptual applications of the introduced formalism. We discuss how negative energies appear in this framework and how to resolve them. That tachyons can acquire negative energies under subluminal Lorentz boosts has already been discussed in \cite{schwartz2018tachyon,paczos2024covariant} and is resolved there by enlarging the Hilbert space. We show here that this problem also affects particles moving at subluminal velocities when they are subjected to superluminal boosts and that it can be resolved by enlarging the Hilbert space as well. We then explicitly calculate the QRF transformations involving an observer in superposition of subluminal and superluminal velocities and argue that from their perspective particles that previously appeared to be outgoing from the perspective of another agent can now appear to be in a superposition (or even entangled with the state of the second agent) of being incoming and outgoing. While previous works have noted this exchange of ``incoming'' and ``outgoing'' for tachyons under subluminal boosts, the case involving boosting subluminal particles superluminally and the appearance of superpositions and entanglement are novel to our work. 
We also show that Bell violations remain invariant under superluminally extended QRF transformations, extending the works of \cite{giacomini2019relativistic,streiter2021relativistic} to our setting, but discuss that their meaning may not remain so. 


\subsection{Deriving superluminal Lorentz transformations}

In this section, we will derive the Lorentz transformations following \cite{parker1969faster}. 
Let us consider the (1+1)-dimensional case, where an inertial frame $(t', x')$ moves with velocity $V$ relative to the frame $(t, x)$. A transformation between these two frames should be linear and its coefficient should depend only on the relative velocity $V$ (the principle of relativity). The inverse of such a transformation should also be given by a sign flip of $V$.  
Hence, such transformations shall be of the form
\begin{gather}
\begin{aligned}
    x'&= A(V)x+ B(V)t,\\
    x&= A(-V)x'+B(-V)t'.
\end{aligned}
\end{gather}
where $A(V)$ and $B(V)$ are unknown functions. Setting $x' = 0$ in the first equation, it follows that these unknown functions are dependent on each other with $B(V)/A(V)=-x/t =-V$. Eliminating $B(V)$ in the above equation and demanding additionally that reversing the direction of the space axis $x$ (which also leads to reversal of the sign of $V$), preserves the form of the transformation equations up to possibly a change of the overall sign, then also tells us that $A(V)$ must either be symmetric or antisymmetric. 
For the symmetric case, $A(-V)=A(V)$, we can retrieve the usual Lorentz transformations (setting $c=1$):

\begin{equation}\label{LB1}
    x'=\frac{x-Vt}{\sqrt{1-V^2}}\\\hspace{1cm}
    t'=\frac{t-Vx}{\sqrt{1-V^2}}
  \end{equation}
For the anti-symmetric case, $A(-V)=-A(V)$, we get the following transformation which is well-behaved for $V>c=1$
\begin{equation}\label{LB2}
    x'=\pm\frac{V}{|V|}\frac{x-Vt}{\sqrt{V^2-1}}\\\hspace{1cm}
    t'=\pm\frac{V}{|V|}\frac{t-Vx}{\sqrt{V^2-1}}.
\end{equation}

The sign in front of these equations cannot be uniquely determined, since there is no limit $V \rightarrow 0$. Hence, the choice of sign is just a matter of convention, and we choose the positive whenever the need arises. Even though there is more than half a century of research concerned with superluminal Lorentz transformations, subtle but at the same time relevant operational aspects of the derivation summarized above, are still under debate\cite{damski2025lorentz}.

\subsection{Quantum reference frames}
In this section, we will briefly review the framework of quantum reference frames. A full account is given in \cite{giacomini2019quantum}.

Reference frames are abstract objects, which are used to specify coordinates and standardise measurements within the reference frame. The laws of physics are the same regardless of the choice of reference frame and physical quantities change covariantly, i.e., according to a representation of the covariance group \cite{ballesteros2021group}. For example, Maxwell equations with sources transform as four-vectors, that is, under the $(1/2,1/2)$ representation of the $O(1,3)$ group. In a laboratory situation, these abstract reference frames can be realised through a physical system which ultimately follows quantum mechanical laws. The description of the quantum state is given in terms of relative quantities w.r.t the chosen reference frame of observation. 

\vspace{0.3 cm}

\begin{figure}[htp]
		\centering
		\includegraphics[width= 0.4 \textwidth]{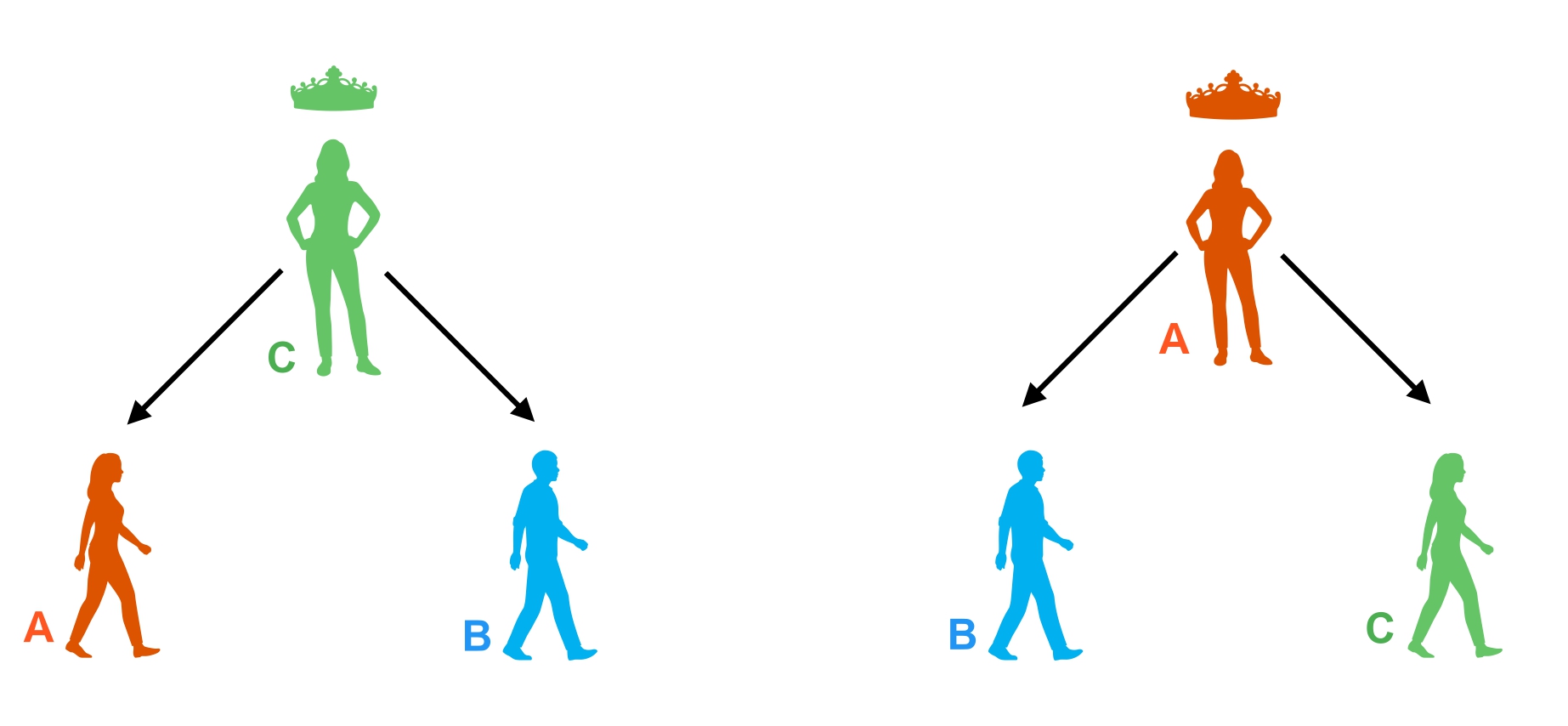}	\caption{From the perspective of reference frame $C$, $A$ and $B$ are the external systems whose degree
of freedom we consider, and similarly from the perspective of $A$, we have two external systems: system $B$ and $C$. This figure is adapted from Fig.2 of \cite{giacomini2019quantum}. }
		\label{f3.1}        
	\end{figure}

\vspace{0.5 cm}

\Cref{f3.1} gives an example with three systems. System $C$ is the initial reference frame, from which we describe the systems $A$ and $B$. We can say that we are in the rest frame of $C$. We then apply a reference frame transformation to go to a new reference frame, that of system $A$, i.e., the rest frame of $A$. The systems $B$ and $C$ are then described from the perspective of $A$. \\

Quantum reference frame transformations allow us then to do such transformations even when the involved systems are quantum. A schematic example is depicted in \cref{f3.1}. Laboratory procedures, such as preparations, transformations, and measurements, have fundamental status, which makes the technique operational by translating into assessable experiments. \\

Reference frame transformations are canonical transformations, which preserve the symplectic structure i.e., the action on the phase space.\footnote{Quantum canonical transformations are not necessarily assumed to be isometries\cite{rovelli1996relational,rovelli2021helgoland, aharonov2008two,vaidman2007two,delahamette2020quantum, carette2025operational,mikusch2021transformation, bojowald2022algebraic,loveridge2018symmetry,lake2023revisited, glowacki2023operational}. However, for our purposes we consider only unitary transformations, which by definition are isometries.} 

\vspace {0.3 cm}



\begin{figure*}[htp]
		\centering
		\includegraphics[width= 1.0 \textwidth]{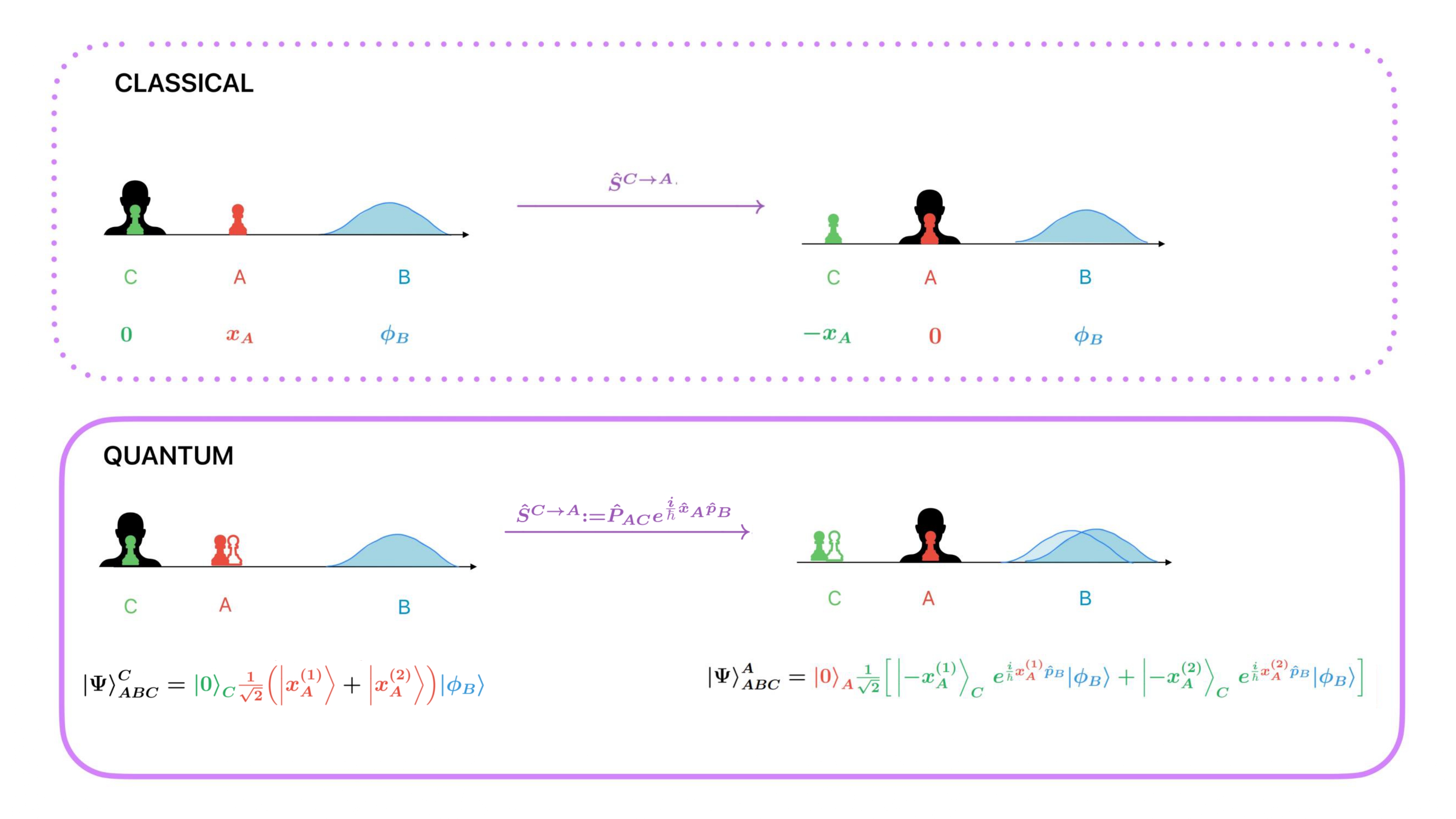}	\caption{Above, \textbf{Classical:}
        Classical reference frame transformation of the quantum state $B$ from $C$ to $A$, essentially relocating the reference frame to the localized position of the new reference frame, denoted by scalar position parameter. The classical reference frame transformation is essentially a Galilean transformation conditioned on the classical position parameter of the new reference frame, here $x_A$ followed by a parity-swap. 
        \\
        Below, \textbf{Quantum:}
        Using the QRF transformation, from $\mathcal{H}^{(C)}_A \otimes \mathcal{H}^{(C)}_B$ to $\mathcal{H}^{(A)}_B \otimes \mathcal{H}^{(A)}_C$, where the superscript denotes the reference frame of the observer, we see that superposition and entanglement are frame-dependent notions \cite{giacomini2019quantum}. In the first figure, from $C$, we see the new reference frame $A$ in superposition, while after the QRF transformation, from the new reference frame $A$, we see our physical system $B$ entangled with the initial reference frame C. This can lead to misconceptions about frame-dependent Bell violations, which we discuss later on. These figures are adapted from Fig.3 of \cite{giacomini2019quantum}. }
		\label{f3.3}        
	\end{figure*}

\begin{figure*}[htp]
		\centering
		\includegraphics[width= 0.8 \textwidth]{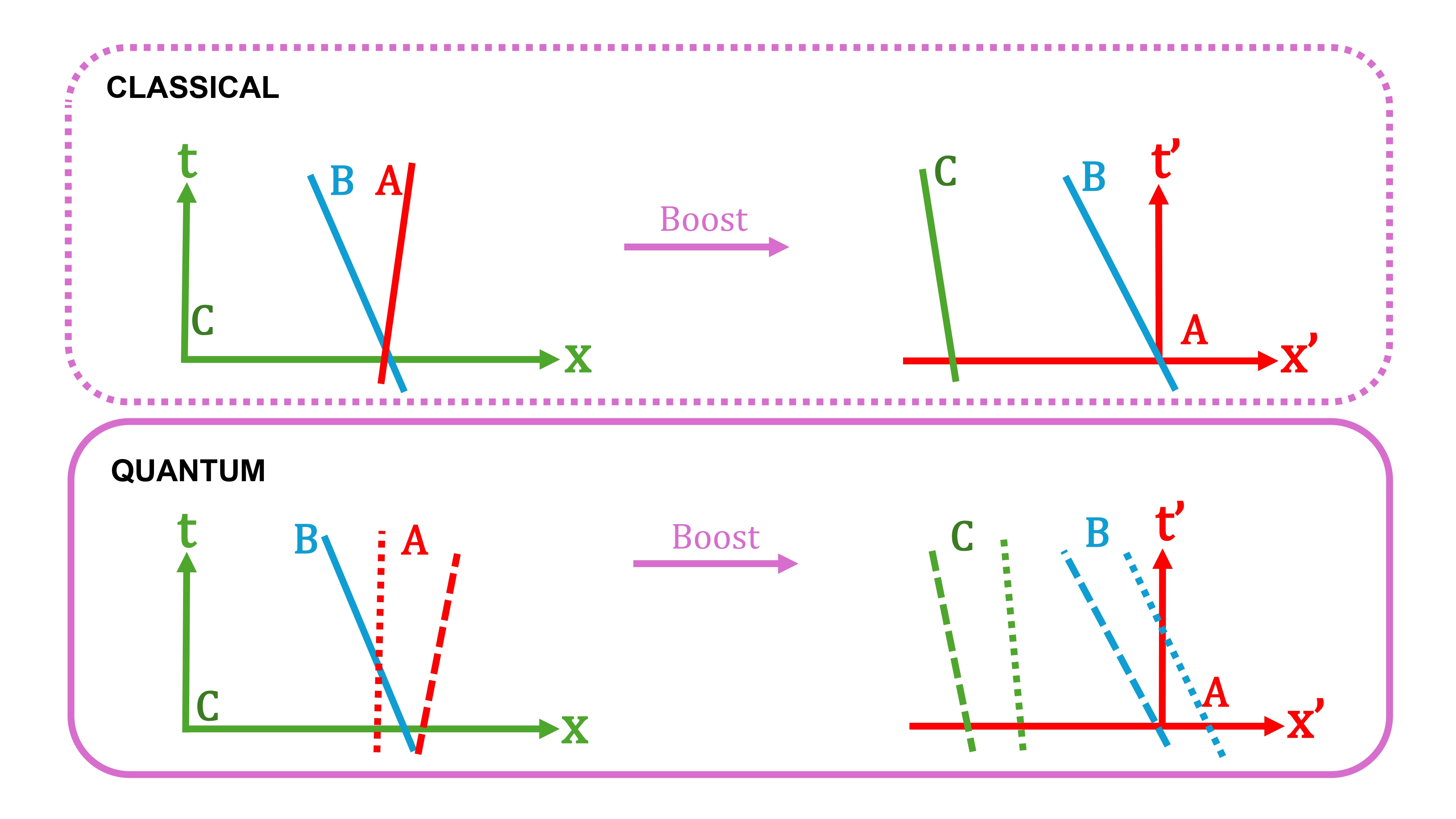}	\caption{Above, \textbf{Classical:}
        Classical reference frame transformation of the world line of $B$ from $C$ to $A$ is essentially relocating the reference frame to the localized position of the new reference frame, denoted by solid world lines. This Lorentz transformation is conditioned on the classical velocity boost of the new reference frame, followed by a parity-swap.   
        \\
        Below, \textbf{Quantum:}
        Using the QRF transformation, from $\mathcal{H}^{(C)}_A \otimes \mathcal{H}^{(C)}_B$ to $\mathcal{H}^{(A)}_B \otimes \mathcal{H}^{(A)}_C$, where the superscript denotes the reference frame of the observer, we see again that superposition and entanglement are frame-dependent notions \cite{giacomini2019quantum}. In the first figure, from $C$, we see the new reference frame $A$ in superposition of different velocities represented by dotted and dashed world lines, while after the QRF transformation, from the new reference frame $A$, we see the physical system $B$ entangled with the initial reference frame $C$. The dotted and dashed delocalised (quantum) world lines of $B$ and $C$ become entangled on the right hand side. Note that the reference frame of observer always reads classical and hence is represented by solid lines.}
		\label{boostQRF}        
	\end{figure*}

Quantum reference frames are physical systems following quantum mechanical laws. This makes it possible to formalize a more generalized reference frame transformation, allowing to transform into a reference frame which is in superposition of measurable parameters \cite{giacomini2019quantum}, using the linearity of quantum theory. \Cref{f3.3} shows the coherent translation of $B$ relative to the position of $A$, via the operator $e^{i\hat{x}_{A}\hat{p}_{B}/\hbar}$  where the indices refer to the quantum systems $A, B$. A quantum essence of this transformation is encoded in replacing the classical parameter of the standard translation operator by the position operator of the system $A$ and similarly for the associated canonical momentum, $x_A \rightarrow \hat{x}_{A}$ and $p_A \rightarrow \hat{p}_A$\footnote{In the QRF transformation, we apply the transformation to one of the conjugate variables, and the complementary variable is then determined from the final transformed state to preserve the uncertainty relations under a particular measurement \cite{giacomini2019quantum}.}. It allows us to apply quantum-controlled (coherent) translations in which the position of the system A becomes the control degree of freedom, possibly in a superposition. Similarly, \Cref{boostQRF} shows the scenario in which the new reference frame $A$ is now in a superposition of Lorentz boosts.

The full spatial translation from $C$ to $A$ is a canonical transformation on the phase space observables of the systems $A$ and $B$ defined by 
\begin{gather}\label{eq:translationqrf}
\begin{aligned}
   &\hat{S}_{AC}: \mathcal{H}_{A}^{(C)} \otimes \mathcal{H}^{(C)}_B \rightarrow \mathcal{H}_{B}^{(A)} \otimes \mathcal{H}^{(A)}_C \\
   &\hat{S}_{AC}= \hat{P}_{AC}e^{i\hat{x}_{A}\hat{p}_{B}/\hbar}
\end{aligned}
\end{gather}
where $\mathcal{H}_{A}^{(C)}$ is the Hilbert space for the state of $A$ from the perspective of $C$, and analogously for the other Hilbert spaces. Note that $\mathcal{H}^{(C)}_B \cong \mathcal{H}^{(A)}_B$ for any choice of systems $A, B, C$. We will sometimes drop the superscript when it is clear whose reference frame we are considering. 
Moreover, the Hilbert space of $C$ does not show up in the domain of $\hat{S}_{AC}$. Conversely, the Hilbert space of $A$ does not show up in its codomain. This is because we are originally in the reference frame of $C$ and do not need to include its external degrees of freedom in the overall description and similarly for $A$ in the end. The so-called parity-swap operator $P_{AC} : \mathcal{H}_{A}^{(C)} \rightarrow \mathcal{H}_{C}^{(A)}$ acts like 
\begin{equation}
    \hat{P}_{AC}\ket{x}_A^C= \ket{-x}_C^A.
\end{equation}
where $\ket{x}_A^C$ denotes the position basis of $\mathcal{H}_A^{(C)}$, and analogously for $\ket{x}_C^A$. Thus, it accounts for the switch of whether $A$ or $C$ is included and represents the exchange of direction of perspective with respect to direction of the boost. On the other hand, $e^{i\hat{x}_{A}\hat{p}_{B}/\hbar}$ describes the coherent translation as already discussed. 

More generally, instead of just translation, we can consider any other kind of canonical reference frame transformation $T_\alpha$. In this case, we replace the translation operator $e^{i x_{A}\hat{p}_{B}/\hbar}$ with a unitary representation 
\begin{equation}
\hat{U}_B(T_{\alpha}): \mathcal{H}^{(C)}_B \rightarrow  \mathcal{H}^{(A)}_B
\end{equation}
of the transformation $T_\alpha$ controlled by some parameter $\alpha$ (note that $e^{i\hat{x}_{A}\hat{p}_{B}/\hbar}$ is obtained by lifting the parameter to an operator in the unitary representation of the group of translations).
 We then obtain the transformation\footnote{The parity represents the exchange of direction of perspective with respect to direction of boost, while the exponential part represents the boosts. Imagine a classical scenario in which an observer $C$ is located at $(0,0)$ and sees $A$ and $B$ located at $(x_1,0)$ and $(x_2,0)$, such that $x_1< x_2$. If now, we want $C$ to swap their position with the position of $A$, we see $C$ at $(-x_1,0)$ while the new location of $A$ is at $(0,0)$. This is exactly the function of the parity-swap operator in general.}
\begin{equation}
    \hat{S}_{AC}= \hat{P}_{AC} \circ\int d\alpha  \ket{\alpha}\bra{\alpha}_A^C \otimes\hat{U}_B(T_\alpha).
\end{equation}
We will also write
\begin{equation}\label{eq:liftedrep}
    \hat{U}_B(T_{\hat{\alpha}}) := \int d\alpha  \ket{\alpha}\bra{\alpha}_A^C \otimes\hat{U}_B(T_\alpha).
\end{equation}

In particular, the formalism of quantum reference frame can describe relativistic settings by considering unitary representations of the Lorentz group \cite{ballesteros2021group}.


  

\section{Superluminal observers in extended QRF framework}

\begin{table*}[htpb]
    \centering
    \renewcommand{\arraystretch}{2.5}
    \begin{tabular}{|c c|}
        \hline
        \textbf{Subluminal} & \textbf{Superluminal} \\
        \hline
        \textbf{Velocity:} \( V \) & \textbf{Dual Velocity:} \( \tilde{V} \) \\
        \( 0 \leq V^2 < 1 \) & \( \tilde{V}^2 = V^{-2}, \quad 0 \leq \tilde{V}^2 < 1, \quad \text{and} \quad \infty > V^2 \geq 1 \) \\
        \hline
        \textbf{Rapidity:} \( \varphi \) & \textbf{Dual Rapidity:} \( \tilde{\varphi} \) \\
        \( V^2 = (\tanh(\varphi))^2 \) & \( \tilde{V}^2 = (\tanh(\tilde{\varphi}))^2 \) \\
        \( 0 \leq \varphi^2 < \infty \) & \( 0 \leq \varphi^2 < \infty, \quad 0 \leq \tilde{\varphi}^2 < \infty \) \\
        \hline
        \(\gamma(x) = \frac{1}{\sqrt{1-V^2}}\) (\( c=1 \)) & \(\tilde{\gamma}(x) = \frac{\tilde{V}}{|V|}\frac{1}{\sqrt{V^2-1}}\) (\( c=1 \)) \\
        \hline
        \( dx' = B_{\varphi}dx = 
        \begin{bmatrix}
        \cosh(\varphi) & \sinh(\varphi) \\
        \sinh(\varphi) & \cosh(\varphi)
        \end{bmatrix} dx
        \) 
        & 
        \( dx' = \tilde{B}_{{\varphi_{\pm}}}dx = \pm
        \begin{bmatrix}
        \sinh(\tilde{\varphi}) & \cosh(\tilde{\varphi}) \\
        \cosh(\tilde{\varphi}) & \sinh(\tilde{\varphi})
        \end{bmatrix} dx
        \) \\
        \hline
        \multicolumn{2}{|c|}{\( B = \{B_{\varphi}, \tilde{B}_{\varphi +}, \tilde{B}_{\varphi -}\} \) forms a group.} \\
        \hline
    \end{tabular}
    \caption{Group of Subluminal and Superluminal Lorentz Transformations \cite{lake2024towards}. The left column represents the subluminal velocity $(V)$ and the parameters: rapidity, Lorentz factor $\gamma(x)$, and the subluminal transformation defined on it. The subluminal rapidity span the whole range of real numbers, leaving the superluminal velocities $(\tilde{V})$ mapped as a shadow of the subluminal velocities on the right hand side of the table. As we can see, the subluminal and superluminal transformation matrices, when re-parameterised with such mapping, the group of subluminal and superluminal Lorentz transformation is closed. }
    \label{box}
\end{table*}

The quantum reference frame framework discussed earlier allows us to include non-classical properties of reference frames like superposition and entanglement of physical frames. The framework assigns quantum probabilities to subluminal physical frames. In this section, we will discuss how we can include superluminal observers by extending this framework.

Recently, \cite{lake2024towards} (and previously \cite{marchildon1983superluminal}), suggested the possibility of embedding the superluminal boosts within a group structure. The authors show that in (1+1) dimensions, while the SpLT do not form a group by themselves, the SbLT and SpLT together form a group $SL(2,\mathbb{R})$ with an asymmetric direct sum. The mapping the authors used for the proof is summarised in \cref{box}. The special linear group $SL(n,\mathbb{R})$ is a Lie group with a well-defined algebraic and topological structure and one can construct unitary representations of $SL(2,\mathbb{R})$ representing the boost in quantum reference frame transformations. The QRF transformations discussed above become
\begin{equation}
    \hat{S}_{AC} = \hat{P}_{AC} \hat{U}_B(L_{\hat{p}_A}),
\end{equation} 
where $\hat{P}_{AC}$ is again the parity-swap operator. 
 The operator $\hat{U}_B(L_{\hat{p}_A})$ is defined with the 2-momentum $p_A$
\begin{equation}
    p_A = \begin{pmatrix}
        p_A^0 \\
        p_A^1
    \end{pmatrix}
\end{equation}
where $p_A^0$ is the energy and $p_A^1$ is the spatial momentum, as in \cref{eq:liftedrep} (in the following, we will keep the superscripts on states indicating the reference frame implicit to reduce clutter)
\begin{gather}
\begin{aligned}
    \hat{U}_B(L_{\hat{p}_A}) = \int& dp_A \left(\delta(p_A^2 - m_A^2) + \delta(p_A^2 + m_A^2)\right) \\
    &\ket{p_A} \bra{p_A}_A \otimes \hat{U}_B(L_{p_A})  \\
    = \int &d\bar{p}_A\ket{p_A} \bra{p_A}_A \otimes \hat{U}_B(L_{p_A})
\end{aligned}
\end{gather}
with $\hat{U}_B(L_{p_A})$ a unitary representation of $SL(2,\mathbb{R})$ representing the boost $L_{p_A}$ associated to 2-momentum $p_A$\footnote{While Lorentz boosts are usually written in terms of the velocity (as we also did in \cref{LB1,LB2}), note that (assuming the mass is known) the 2-momentum uniquely determines the velocity.} acting as
\begin{equation}
    \hat{U}_B(L_{p_A}) \ket{p_B}_B = \ket{L_{p_A} p_B}_B.
\end{equation}
Here, $L_{p_A} p_B$ is the 2-momentum after applying the boost $L_{p_A}$ (i.e., boosting with the velocity which corresponds to 2-momentum $p_A$ given mass $m_A$). 

The first delta function represents the subluminal momenta while the second one represents the superluminal momenta. The measure $d\bar{p}_A = dp_A \left(\delta(p_A^2 - m_A^2) + \delta(p_A^2 + m_A^2)\right)$ with $dp_A=dp_A^0 dp_A^1$ and $p_A^2 = (p_A^0)^2 - (p^1_A)^2$ is then the 2-dimensional Lorentz invariant integration measure. This is indeed Lorentz invariant under both superluminal and subluminal boosts. Consider a change of variables $p_A \rightarrow p'_A = Lp_A$ with $L$ a subluminal or superluminal Lorentz boost. Then,
\begin{equation}
    d\bar{p}_A = dp'_A |\text{det}{L}|^{-1} (\delta((L^{-1} p'_A)^2 - m_A^2) + \delta((L^{-1} p'_A)^2 + m_A^2).
\end{equation}
Note now that $\det{L} = 1$ when $L$ is a subluminal boost and $\det{L} = -1$, when it is a superluminal boost. Further, for a subluminal boost $L$, we have that $(L^{-1} p'_A)^2 = p^{\prime 2}_A$ while for a superluminal boost, we have $(L^{-1} p'_A)^2 = -p^{\prime 2}_A$. Thus, in the latter case, we have
\begin{gather}
\begin{aligned}
    \delta((L^{-1} p'_A)^2 &- m_A^2) + \delta((L^{-1} p'_A)^2 - m_A^2 \\&= \delta(-p_A^{\prime 2}  - m_A^2) + \delta(- p^{\prime 2}_A + m_A^2) \\
    &= \delta(p_A^{\prime 2}  + m_A^2) + \delta( p^{\prime 2}_A - m_A^2)
\end{aligned}
\end{gather}
where we used that $\delta(-x) = \delta(x)$. For both superluminal and subluminal boosts, we then find
\begin{equation}
    d\bar{p}_A = d\bar{p}'_A.
\end{equation}

The action on the states in terms of the momentum basis\footnote{One could also work in the position basis. This could be done by doing the transformation in the momentum basis and then applying Fourier transforms. Alternatively, in particular if all one cares is about $B$ being written in terms of position, the transformation is given by $\hat{S}_{AC}\ket{p_{A}}_{A}  \ket{x_B}_{B} = \ket{m_A^{-1} m_C p_A^1|p_A^1|^{-1} (p_A^0, -p^1_A)}_{C} \ket{L_{p_A} x_B}_{B}$ where $x_B$ is the initial spacetime coordinate of $B$ and $L_{p_A} x_B$ is the corresponding spacetime coordinate after a boost corresponding to momentum $p_A$.} for a superluminal momentum $p_A$ is then given by 
\begin{gather}
\begin{aligned}
    \hat{S}_{AC}&\ket{p_{A}}_{A}\ket{p_B}_{B}  \\
    &= P_{AC} \circ \int d\bar{p}'_A \ket{p'_A}_A \braket{p'_A|p_A}_A \otimes \hat{U}_B(L_{p'_A}) \ket{p_B}_B\\
    &=\ket{m_A^{-1} m_C p_A^1|p_A^1|^{-1} (p_A^0, -p^1_A)}_{C} \ket{L_{p_A} p_B}_{B}
\end{aligned}
\end{gather}
where $L_{p_A} p_B$ is the action of a Lorentz boost corresponding to the momentum $p_A$ on the initial 2-momentum $p_B$ of particle $B$. The masses $m_C$ and $m_A$ are the masses of particle $C$ and particle $A$. These appear in the equation above because the state of $C$ in the reference frame of $A$ corresponds to boosting a state at rest with the momentum $p_A$, thus,
\begin{gather}
\begin{aligned}
    p'_C = L_{p_A} \begin{pmatrix}
        m_C \\
        0
    \end{pmatrix} &= \frac{p_A^1}{|p_A^1|} \frac{m_C}{\sqrt{v_A^2 - 1}} \begin{pmatrix}
        1 \\
        -v_A
    \end{pmatrix}\\
    &=\frac{p_A^1}{|p_A^1|} \frac{m_C}{m_A} \begin{pmatrix}
        p_A^0 \\
        -p_A^1
    \end{pmatrix}
\end{aligned}
\end{gather}
where $v_A$ is the velocity of $A$ in the reference frame of $C$.\footnote{The reader might notice that in the above the energy takes on negative values when $p_A^1 < 0$. In fact, this appearance of negative energies is a broader issue which we shall resolve in \cref{sec:energy}.} We can also use this to write the action of the parity-swap operator in terms of the momentum basis
\begin{equation}
    \hat{P}_{AC} \ket{p_A}_A = \ket{m_A^{-1} m_C p_A^1|p_A^1|^{-1} (p_A^0, -p^1_A)}_{C}
\end{equation}
Now, the unitary transformation on a closed group retains the structure of the group and closes on itself. This follows by definition from the fact that we are dealing with a unitary representation. Therefore, since $SL(n,\mathbb{R})$ is closed, the unitary transformations on it representing the quantum reference frame transformation also form a closed group. This observation concludes our construction.

\section{Application of the extended QRF framework}

In this section, we discuss how the framework of extended QRF can be used to address physically relevant issues that naturally arise with superluminal observers. In particular, we look at apparent paradoxes involving the energy, entropy and temperature under superluminal Lorentz boosts and show how to resolve them. We further show that Bell violations are invariant in the extended QRF just as they are in the subluminal setting. 

\subsection{Energy under superluminal boosts}\label{sec:energy}
We are going to analyze the energy of particles under superluminal (and subluminal) QRF transformations. In particular, we will see that energies can appear to become negative. This problem has been noted and resolved before for the case of tachyons being transformed subluminally in \cite{schwartz2018tachyon} and \cite{paczos2024covariant}. 

\begin{figure}[htp]
		\centering
		\includegraphics[width= 0.5 \textwidth]{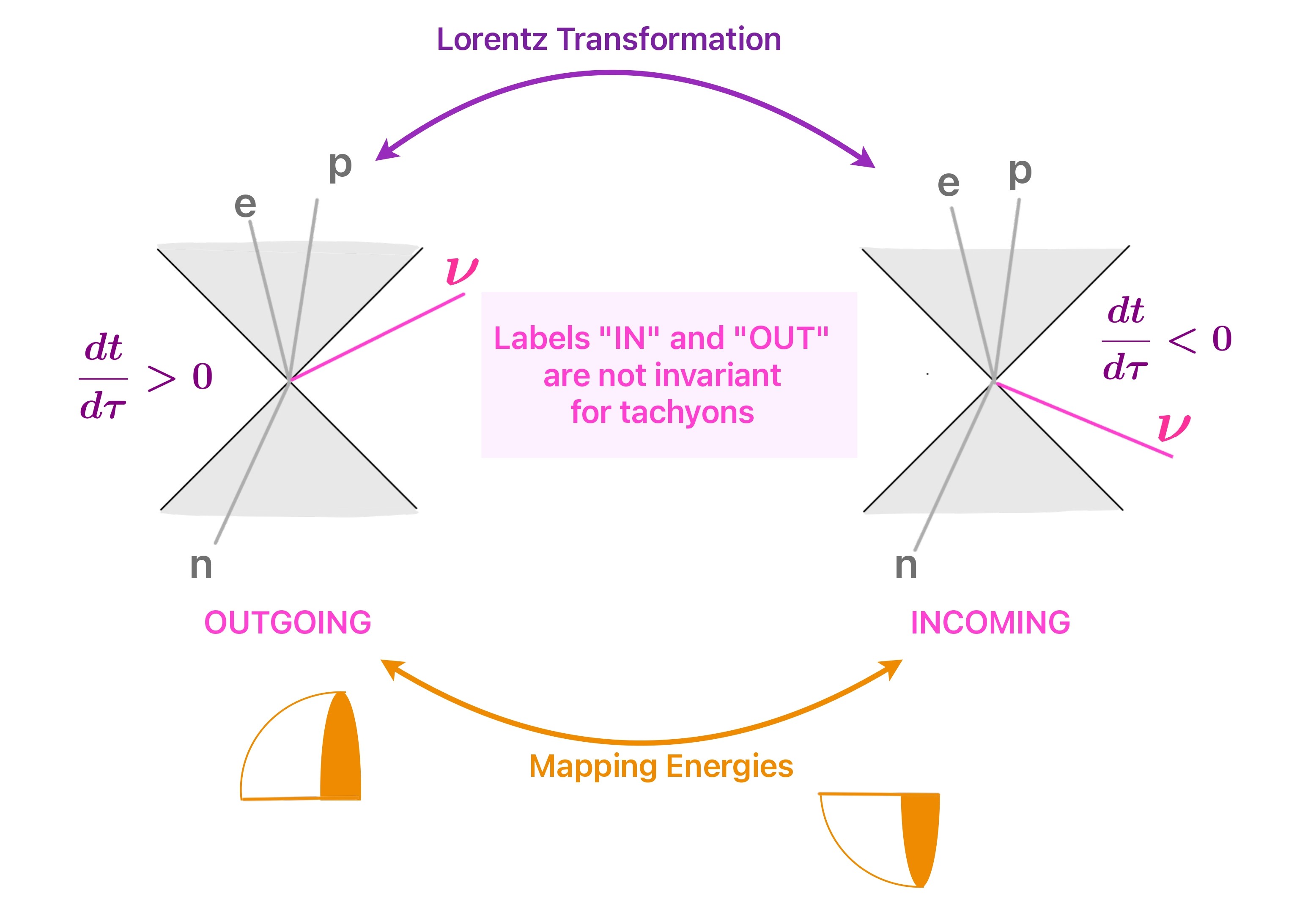}	\caption{The space of single particle states is not Lorentz invariant, hence should be enlarged. $e, p, n,$ and $\nu$ correspond to electron, proton, neutron, and superluminal tachyon, respectively. The direction and velocity of the tachyon is decided by the differential ratio of coordinate time: $t$ and proper time: $\tau$ of the tachyon in each reference frame differed by a subluminal Lorentz transformation. The figure is adapted originally from figure 1 of\cite{schwartz2018tachyon} and similar adaptations are also used more recently in \cite{paczos2024covariant}.
}
		\label{f3.6}        
	\end{figure}

We illustrate a scenario analysed in \cite{schwartz2018tachyon} in \cref{f3.6}. On the left hand side a neutron decays into a proton, an electron and a neutrino. The neutrino here is a tachyon and all particles have positive energy. However, under a (subluminal) Lorentz boost, depicted on the right hand side of \cref{f3.6}, the tachyonic neutrino acquires a negative energy. The solution that \cite{schwartz2018tachyon} proposes is to reinterpret the particle from an outgoing to an incoming one. The naïve state space of a particle is, hence, not Lorentz invariant, but can be made so by expanding it.

\subsubsection{Energy accounting}    

\begin{figure}[htp]
		\centering
		\includegraphics[width= 0.5 \textwidth]{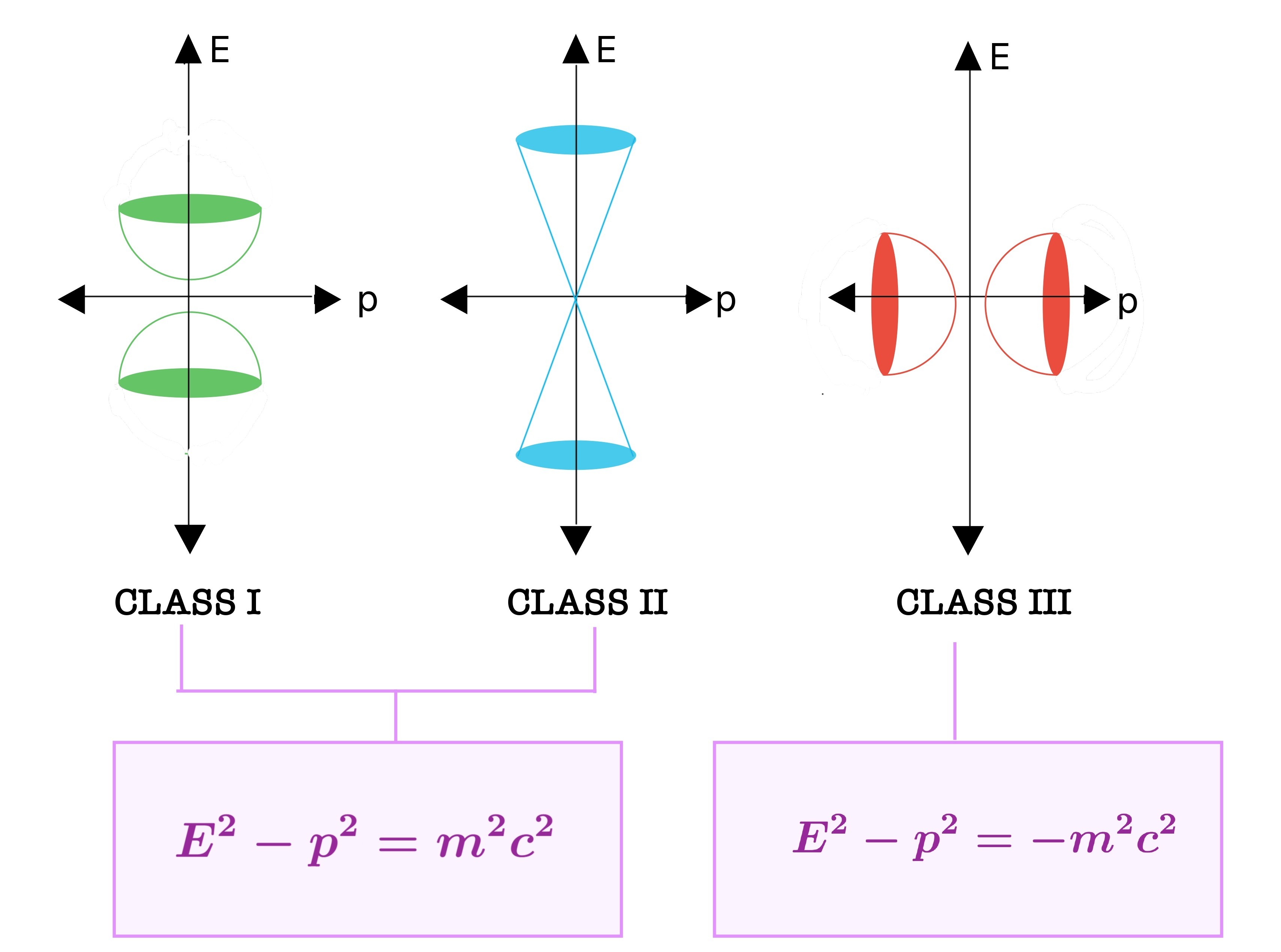}	\caption{Energy-momentum relations for class I: subluminal particles, class II: photons, class III: tachyons}
		\label{f3.4}        
	\end{figure}
The energy-momentum relations derived from the postulates of special relativity can be broadly classified into two categories: class I, the subluminal particles, i.e., the usual physical systems we encounter in nature and class II: massless particles traveling at the speed of light, e.g., photons. Finally, by including superluminal Lorentz transformations, we obtain class III corresponding to superluminal particles, i.e., tachyons. We depict their energy-momentum relations in \cref{f3.4}. 

The regions on the continuous surface of the hyperboloids (cone in the case of photons) are transformable within each other under subluminal Lorentz transformations, while superluminal Lorentz transformations allow us to transform within one class or switch classes between I and III (however, not from or to class II as the speed of light is constant in all reference frames). This means that class III particles which have positive energies in one reference frame can be transformed with a subluminal Lorentz transformation to a new reference frame where they have negative energies as discussed in the previous section already. At the same time, class I particles can be transformed with a superluminal Lorentz transformation into class III particles and thus can also acquire negative energies while class II particles can be transformed to the negative energy cone by a superluminal transformation.

Additionally, this so far classical situation can be extrapolated to quantum superpositions by allowing our physical reference frames to be in superposition of subluminal/superluminal velocities. 

Let us now look at the precise energy relations for subluminal and superluminal Lorentz transformations. The Lorentz-boosted energy equations have the form 
\begin{equation}\label{Eprime}
E'= \gamma (E-Vp)
\end{equation}
where $E$ is the energy of $B$ in the previous reference frame, $E'$ is the energy in the Lorentz-boosted reference frame, $V$ is the relative velocity of the two frames (i.e., is the relative velocity of $A$ w.r.t $C$), and $p$ is the momentum of $B$ in the original reference frame (which we choose to be positive w.l.o.g.). We will assume all quantities to have sharp values as the case of superpositions follows from linearity. At this point, let us acquire the notation of \cite{lake2024towards} where we write $V$ for velocities $0<|V|<1$ (setting $c=1$) and $\tilde{V}$ for velocities $1<|\tilde{V}|< \infty$.  







We assume that the particle $B$ is initially \textit{subluminal} or a \textit{photon} in the reference frame $C$. If $A$ is also subluminal w.r.t to $C$, then the Lorentz factor is of the form $\gamma= \frac{1}{\sqrt{1-V^2}}$. Hence, to obtain positive energy in \cref{Eprime}, we need 
\begin{equation}
    E/p >V \text{ or } V<1< \sqrt{\frac{m^2}{p^2}+1}
\end{equation}
where we used that $E/p = \sqrt{\frac{m^2}{p^2}+1}$ (note that this holds for photons with $m=0$ as well). We note that these conditions hold even when $V < 0$, hence negative energies are not possible in this case.

On the other hand, if $A$ is superluminal w.r.t to $C$, the Lorentz factor is of the form $\tilde{\gamma}= \frac{\tilde{V}}{|\tilde{V}|}\frac{1}{\sqrt{\tilde{V}^2-1}}$. Hence, to get positive energy with positive velocity $\tilde{V} > 0$, the boost velocity needs to satisfy
\begin{equation}\label{condsuppos}
    \sqrt{\frac{m^2}{p^2}+1}> \tilde{V}
\end{equation}. 





To get positive energy with negative velocity $\tilde{V} < 0$, the boost velocity needs to satisfy 
\begin{equation}
    \tilde{V}> \sqrt{\frac{m^2}{p^2}+1}.
\end{equation}
Since the velocity is negative in this case and the r.h.s. is always positive, positive energy is unachievable.

Turning the above two conditions around, to get negative energy with positive velocity, the boost velocity needs to satisfy 
\begin{equation}
    \tilde{V}> \sqrt{\frac{m^2}{p^2}+1}.
\end{equation} 

To get negative energy with negative velocity, the boost velocity needs to satisfy 
\begin{equation}
    \sqrt{\frac{m^2}{p^2}+1}> \tilde{V},
\end{equation} 
which is always true. We notice that photons always acquire a negative energy under a superluminal boost. A concise summary of the above conditions is presented in Table II.\\
\begin{table*}[htpb]
\centering
\renewcommand{\arraystretch}{2.5}
\begin{tabular}{|p{5.2cm}|c|p{5cm}|}
\hline
\hspace{1.3cm} \textbf{Applied boost} & \textbf{Negative energy regime} & \hspace{1.4cm}  \textbf{Explanation} \\ \hline

Subluminal boost ($|V|<1$) &
$\displaystyle \sqrt{\frac{m^2}{p^2}+1} < V$ & Impossible as $V < 1 < \sqrt{\frac{m^2}{p^2}+1}$. \\ \hline

Superluminal boost with positive velocity $\tilde V>0$ &
$\displaystyle \sqrt{\frac{m^2}{p^2}+1} < \tilde V$ &
Possible, for large enough boosts as $\tilde{V} > 1$ \\ \hline


Superluminal boost with negative velocity $\tilde V<0$ &
$\displaystyle \tilde V < \sqrt{\frac{m^2}{p^2}+1}$ &
Always satisfies since $\tilde{V} < 0$. Positive energy is impossible for negative superluminal boosts. \\ \hline

\end{tabular}
\caption{Analysis of energy transformation of an initially subluminal particle under subluminal and superluminal boosts.}
\label{tab:superluminal_energy}
\end{table*}
\\


We can take care of negative energies by simply reinterpreting the particle's energy and momentum with

\begin{equation}\label{reinterpet}
    (E, p) \rightarrow (-E, -p),
\end{equation}
extending the solution of \cite{schwartz2018tachyon,paczos2024covariant} to the case of superluminal boosts. This is analogous to reinterpreting a negative energy ``incoming'' particle as a positive energy ``outgoing'' particle (or vice versa).

\subsubsection{Resolution within superluminal QRF}
For our quantum reference frames in order to account for this issue while retaining unitarity, this implies that the state space of each particle is actually larger than if we considered only subluminal transformations. We can resolve this by adapting the solution proposed in \cite{schwartz2018tachyon,paczos2024covariant}.

For each ``positive energy'' basis state $\ket{p_B, \Sigma(b)}_{S_B B}$ (i.e., particle with definite 2-momentum $p_B$, and spin $\Sigma(b)$ with momentum space denoted by $B$ and space for spin denoted by $S_B$\footnote{We have introduced spin (discussed more in the next section), because at relativistic speed, spin and momentum couples with each other which needs to be included in the definition of state.}), there has to exist a ``negative energy'' state $\ket{-p_B, \Sigma(b)}_{S_B B}$. Of course, $-p_B$ is ultimately just a label of the state so there is nothing stopping us from reinterpreting this ``negative energy'' state as a state with positive energy and 2-momentum $p_B$. To make this more apparent, we can split our space into two parts via the reinterpretation

\begin{gather}
\begin{aligned}
    \ket{p_B, \Sigma(b)}_{S_B B} &\cong \ket{p_B, \Sigma(b)}_{S_B B} \\
    \ket{-p_B, \Sigma(b)}_{S_B B} &\cong \ket{p_B, \Sigma(b)}_{S_{B^*} B^*}
\end{aligned}
\end{gather}
where $B^*$ is a copy of $B$. We can interpret $B$ as corresponding to ``incoming'' particles and $B^*$ as ``outgoing'' particles, in line with the ideas discussed at the beginning of this section. Note that we cannot simply map $\ket{-p_B, \Sigma(b)}_{S_B B}$ into $\ket{p_B, \Sigma(b)}_{S_B B}$ as this would destroy the unitarity of our QRF transformations.

We could also think of whether a state belongs to $B$ or $B^*$ as an internal degree of freedom, writing
\begin{gather}
\begin{aligned}
    \ket{p_B, \Sigma(b)}_{S_B B} &\cong \ket{p_B, \Sigma(b)}_{S_B B} \ket{in}_{B'} \\
    \ket{p_B, \Sigma(b)}_{S_{B^*} B^*} &\cong \ket{p_B, \Sigma(b)}_{S_{B} B} \ket{out}_{B'}.
\end{aligned}
\end{gather}
Which formulation to use is of course a matter of preference. Using a separate system, modelling an internal degree of freedom, would be more in line with how previous literature on QRFs would handle it while using the dual space formulation is more in line with works on tachyons. 

\subsection{What will we see from Superluminal Quantum Reference Frames?}\label{sec:supsee}

In this section, we will take a more detailed look at what a superluminal observer sees from their frame of reference. Or, perhaps more interesting, what does an observer who is in a superposition of subluminal and superluminal velocities see.

We consider an agent $C$, the initial frame of reference, who sends out a photon with the momentum $p_B$. The observer $A$ is in superposition of a subluminal momentum $p_A$ and a superluminal momentum $\tilde{p}_A$. We will take the spatial parts of both of these momenta to be positive. We can write this state as
\begin{equation}
    \frac{1}{\sqrt{2}}(\ket{p_A}_A + \ket{\tilde{p}_A}_A) \ket{p_B}_{B^*} \ket{(m_C, 0)}_{C}
\end{equation}
where we chose to place the photon into the dual space to indicate that it is outgoing, instead of being incoming.

Applying the QRF transformation to the rest frame of $A$ is straightforward and yields
\begin{gather}
\begin{aligned}
    \frac{1}{\sqrt{2}}&\ket{(m_A, 0)}_A \left(\ket{L_{p_A} p_B}_{B^*} \ket{m_C/m_A(p_A^0, -p_A^1)}_{C}\right. \\
    &\left.+ \ket{L_{\tilde{p}_A} p_B}_{B^*} \ket{m_C/m_A(\tilde{p}_A^0, -\tilde{p}_A^1)}_{C}\right).
\end{aligned}
\end{gather}
Since $B$ is a photon, under the superluminal boost associated with $\tilde{p}_A$ the energy of the photon will become negative. Thus, we find that the state $\ket{L_{\tilde{p}_A} p_B}_{B^*}$ actually lives in the dual space (the dual space of the dual space is of course the original space). Writing this explicitly yields
\begin{gather}
\begin{aligned}
    \frac{1}{\sqrt{2}}&\ket{(m_A, 0)}_A (\ket{L_{p_A} p_B}_{B^*} \ket{m_C/m_A(p_A^0, -p_A^1)}_{C} \\
    &+ \ket{-L_{\tilde{p}_A} p_B}_{B} \ket{m_C/m_A(\tilde{p}_A^0, -\tilde{p}_A^1)}_{C}).
\end{aligned}
\end{gather}
We can interpret this state as follows: from the perspective of $A$, it is no longer certain that the photon is outgoing. Rather, $A$ perceives a superposition between a subluminal agent with an outgoing photon and a superluminal agent with an incoming photon. This is similar to the case discussed by \cite{schwartz2018tachyon} where an outgoing tachyon becomes an incoming tachyon. The novelty here, in addition to the fact that we can observe this behaviour even with a photon, is that we do not simply switch between the two ``classical'' states ``incoming'' and ``outgoing'' but observe them in superposition and indeed they can even become entangled with other systems.

We can complicate the above idea in an interesting way by having $C$ send a Gaussian wave packet instead of a photon with a fixed momentum. For this purpose, it will be easier to work in terms of rapidities $\xi$, in which case we can write a (subluminal) Gaussian wave packet centered around the rapidity $\xi_0$ and with variance $\sigma^2$ as
\begin{equation}
    \ket{\Phi}_B^C = \int_{-\infty}^{\infty} \frac{d\xi}{2} \mathcal{N} \exp(-\frac{(\xi - \xi_0)^2}{4\sigma^2}) \ket{\xi}_{B^*}
\end{equation}
where $\ket{\xi}_{B^*} = \ket{(m_B \cosh \xi, m_B \sinh \xi)}_{B^*}$ and $\mathcal{N}$ is a normalisation constant.\\ 

Boosting a state $\ket{\xi}_B$ by a subluminal rapidity $\varphi$, yields 
\begin{equation}
    \hat{U}_B(L_\varphi)\ket{\xi}_{B^*} = \ket{\xi-\varphi}_{B^*}
\end{equation}
which can be straightforwardly calculated using the form of the Lorentz boost in terms of the rapidity (see \cref{box} for the definition of the rapidity for subluminal and superluminal velocities). For a superluminal rapidity $\tilde{\varphi} > 0$, we find (notice the exchange of roles of $\sinh$ and $\cosh$ compared to the subluminal $\ket{\xi}_{B^*}$)
\begin{equation}
    \hat{U}_{B^*}(L_{\tilde{\varphi}})\ket{\xi}_{B^*} = \ket{m_B\begin{pmatrix}
\sinh (\xi - \tilde{\varphi})\\
\cosh (\xi - \tilde{\varphi})
\end{pmatrix}}_{B^*}
\end{equation}
Note that the energy is negative whenever $\xi - \tilde{\varphi} < 0$ in which case we should again apply our insights from \cref{sec:energy},
\begin{equation}
    \ket{m_B\begin{pmatrix}
\sinh (\xi - \tilde{\varphi})\\
\cosh (\xi - \tilde{\varphi})\end{pmatrix}}_{B^*} \mapsto \ket{-m_B\begin{pmatrix}
\sinh (\xi - \tilde{\varphi})\\
\cosh (\xi - \tilde{\varphi})\end{pmatrix}}_{B}.
\end{equation}
We will write the above states more compactly with $\xi > 0$
\begin{gather}
\begin{aligned}
    \ket{\xi}_{B(sup)} &= \ket{(m_B \sinh \xi, -m_B\cosh\xi)}_B \\
    \ket{\xi}_{B^*(sup)} &= \ket{(m_B \sinh \xi, m_B\cosh\xi)}_{B^*}
\end{aligned}
\end{gather}
We assume now that the system $A$ is in a superposition $\frac{1}{\sqrt{2}}(\ket{-\varphi}_A + \ket{-\tilde{\varphi}}_A)$. Once again the final state corresponds to a conditional boost acting on the degrees of freedom of the new reference frame. Therefore, after performing the quantum reference frame transformation the resulting state from the point of view of $A$ takes the following form:
\begin{equation}
\begin{split}
&\ket{\varphi}_{C({\text{sub}})}
\int_{-\infty}^{\infty} 
\frac{d\xi}{2}\, 
\mathcal{N} 
\exp\!\left[-\frac{(\xi - (\xi_0-\varphi))^2}{4\sigma^2} \right]
\ket{\xi}_B
\\[6pt]
& + 
\ket{\tilde{\varphi}}_{C({\text{sup}})}
\int_{-\infty}^{0} 
\frac{d\xi}{2}\, 
\mathcal{N} 
\exp\!\left[-\frac{(\xi - (\xi_0-\tilde{\varphi}))^2}{4\sigma^2} \right]
\ket{|\xi|}_{B(sup)}
\\[6pt]
& + 
\ket{\tilde{\varphi}}_{C({\text{sup}})}
\int_{0}^{\infty} 
\frac{d\xi}{2}\, 
\mathcal{N} 
\exp\!\left[-\frac{(\xi - (\xi_0-\tilde{\varphi}))^2}{4\sigma^2} \right]
\ket{\xi}_{B^*(sup)}
\end{split}
\end{equation}
where $\ket{\varphi}_{C(sub)}= \ket{m_C\begin{pmatrix}
\cosh \varphi\\
\sinh \varphi
\end{pmatrix}}_{C(sub)}$ corresponds to a subluminal momentum while $\ket{\tilde{\varphi}}_{C(sup)}=\ket{m_C\begin{pmatrix}
\sinh \tilde{\varphi}\\
\cosh \tilde{\varphi}
\end{pmatrix}}_{C(sup)}$\\
corresponds to a superluminal momentum. 

The signal corresponding to the subluminal part of the superposition is, unsurprisingly, still a Gaussian, simply shifted by the boost rapidity $\varphi$. The superluminal part is more interesting. One can still think of it as a Gaussian if one interprets $\ket{|\xi|}_{B(sup)}$ as the negative momenta counterparts of $\ket{\xi}_{B^*(sup)}$. However, at the same time the wave packet now does not simply consist of left-moving and right-moving plane waves, one of which is exponentially suppressed compared to the other. Rather it consists of parts that are outgoing (corresponding to $B^*(sup)$) and parts that are incoming (corresponding to $B(sup)$), where again one of them is exponentially suppressed compared to the other. 

Due to the exponential suppression, we might wish to say that the wave packet is outgoing if the sign of $\xi_0 - \tilde{\varphi}$ is positive and incoming if the sign is negative (analogous to how we distinguish between left- and right-moving wave packets). This does not return us to the simple ``classical'' picture, however. Let us imagine that $A$ were to measure whether the wave packet is incoming or outgoing and let us ignore the part of the superposition corresponding to $C$ moving at subluminal velocities. In cases where $\xi_0 - \tilde{\varphi}$ is significantly different from 0, we would expect to find either the outcome ``incoming'' or ``outgoing'' depending on the sign of $\xi_0 - \tilde{\varphi}$ with essentially unit probability. However, if $|\xi_0 - \tilde{\varphi}| \approx \sigma$, then it is reasonably likely to obtain either outcome.




\subsection{Entropy in the expanded space}

The second law of thermodynamics dictates that entropy increases in the direction of time. However, denoting the direction of time is a bit tricky when transforming between relativistic reference frames. 

\begin{figure}[htp]
		\centering
		\includegraphics[width= 0.5 \textwidth]{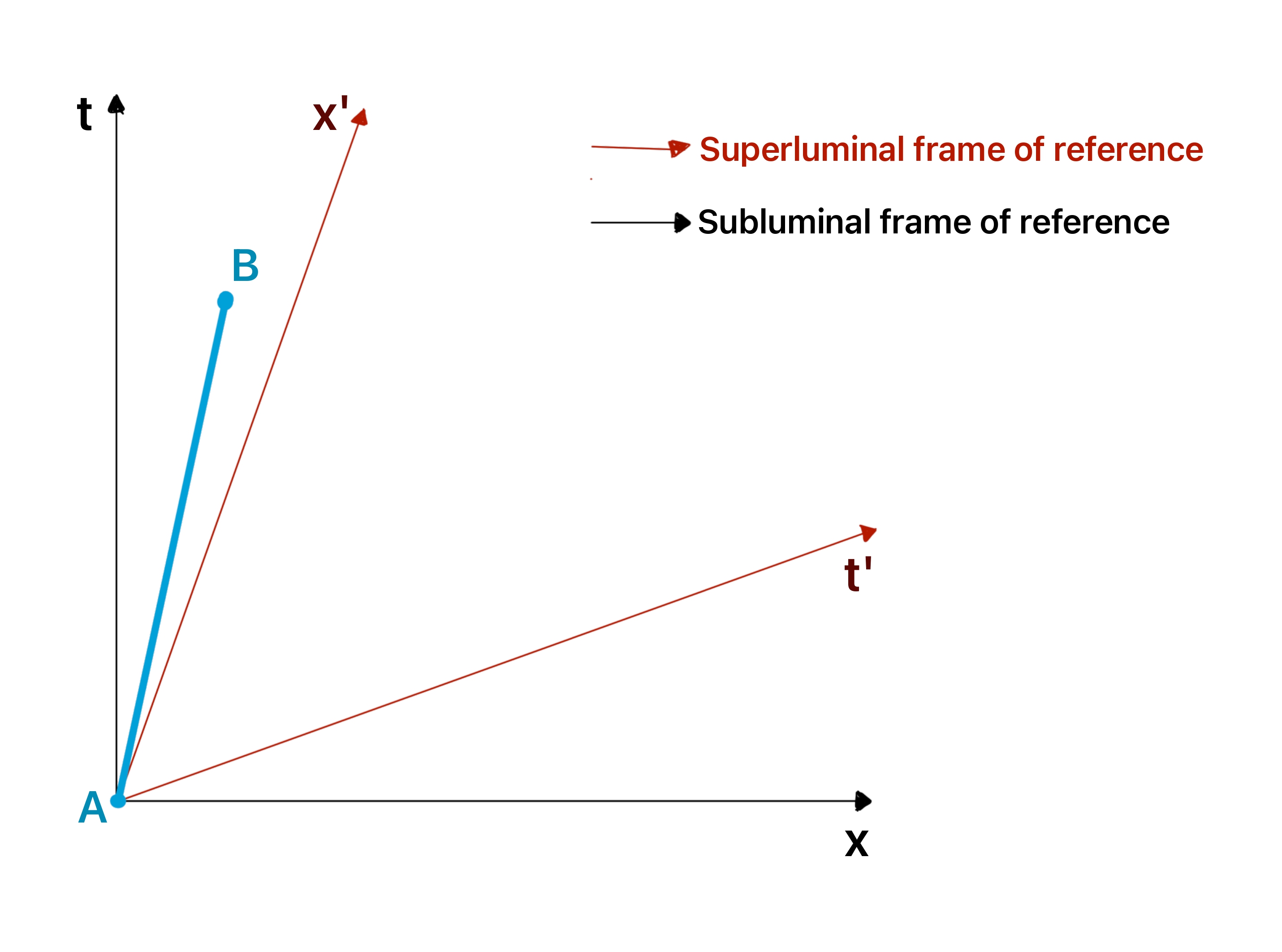}	\caption{Frame dependent notion of time. Point $B$ is located at a positive time coordinate with respect to the subluminal reference frame, but point $B$ is located at a negative time coordinate with respect to the superluminal reference frame. A process that takes any particle from the space-time location $A$ to the space-time location $B$ in the subluminal reference frame will look like it takes the particle from the space-time location $B$ to the space-time location $A$ in the superluminal reference frame. }
		\label{f3.5}        
	\end{figure}

Consider the spacetime points $A$ and $B$ in \cref{f3.5}. In the subluminal un-primed reference frame the direction of time is from  $A$ to $B$ whereas the direction of time is from $B$ to  $A$ in the superluminal primed frame of reference. The relative order of events is a frame-dependent notion even for subluminally boosted reference frames, but only for space-like separated events. This all the more brings us to discuss the definitions of entropy in relativistic and particularly super-relativistic reference frames and check for consistency of the second law of thermodynamics in the presence of superluminal observers.


However, in general there is no consensus on the transformation of thermodynamic quantities under Galilean or Lorentz boosts. An in-depth review can be found in \cite{farias2017temperature}. In short, there are four different approaches stemming from different assumptions and leading to different transformation laws for temperature.
\begin{enumerate}
\item The first approach by Einstein and Planck \cite{einstein1907uber,planck1908dynamik} assumes entropy to be a Lorentz invariant, and thus one finds:
\begin{eqnarray}
dS'=dS,\quad dQ'=\frac{dQ}{\gamma}\quad T'=\frac{T}{\gamma}
\end{eqnarray}
where $\gamma$\footnote{Note: The absolute value of $\gamma$ in the definitions we use, increases as we approach the speed of light both from the subluminal and superluminal regime. This can also be understood from the dual velocity relations discussed in Table I. } is the Lorentz factor, and thus objects look cooler to a (subluminally) moving observer. 

In a superluminal setting, the entropy will remain invariant, however, since $\tilde{\gamma}$ can be negative, we can obtain a negative temperature, which is a problem. We can view this as a consequence of the negative energy problem discussed in the previous section. If the velocities of the particles making up the system are low compared to the speed of light, then to very good approximation the condition from \cref{sec:energy} will be satisfied if and only if the boost velocity $\tilde{V}$ is negative. Hence, whenever this is the case, we should reinterpret the particles' energies, flipping the sign of the change in heat and the temperature. Since the sign of $\tilde{\gamma}$ is the sign of $\tilde{V}$, we find the reinterpreted quantities

\begin{eqnarray}
dS'=dS,\quad dQ'=\frac{dQ}{|\tilde{\gamma}|}\quad T'=\frac{T}{|\tilde{\gamma}|}.
\end{eqnarray}




\item The second approach by Ott \cite{ott1963lorentz} treats heat transfer as an energy, and thus:
\begin{eqnarray}
dQ'=\gamma dQ,
\end{eqnarray}
which, together with the assumption of entropy being a Lorentz invariant, leads to:
\begin{equation}
    T'=\gamma T
\end{equation}
The same argument applies here and we obtain the reinterpreted quantities

\begin{eqnarray}
dS'=dS,\quad dQ'=|\tilde{\gamma}| dQ\quad T'=|\tilde{\gamma}|T.
\end{eqnarray}


\item The third approach by Lansberg \cite{landsberg1966does} directly assumes the temperature to be a Lorentz invariant: $T'=T$.
In order to ensure this one has to define the temperature as:

\begin{equation}    
\frac{1}{T}=\frac{1}{\gamma}\left(\frac{\partial S}{\partial E}\right)_{V,p}
\end{equation}
which also implies that the internal energy is a Lorentz invariant, so that;

\begin{equation}    
\frac{1}{T}=\left(\frac{\partial S}{\partial U}\right)_{V,p}
\end{equation}
With the temperature being constant, the entropy and energy have similar transformations, $\partial S= \gamma \partial E$. Hence, in the expanded space, using reinterpretation, there in no violation of second law as well. 

\item The fourth and final approach by Cavalleri and Salgarelli \cite{cavalleri1969revision} states that it only makes sense to study thermodynamics in the rest reference frame, hence this case is trivial. 
\end{enumerate}



\subsection{Bell violations in the superluminal regime}\label{sec:bell}


For relativistic particles, quantum field theory predicts that the total angular momentum is conserved instead of the spin alone. The spin gets entangled with the momentum in Lorentz boosted reference frames, and for a particle moving in a superposition of velocities, it is impossible to ``jump'' to its rest frame, where the spin is unambiguously defined. One relatively recent paper \cite{giacomini2019relativistic} provides the operational procedure with a QRF transformation corresponding to a ``superposition of Lorentz boosts'', allowing us to transform to the rest frame of a particle that is in a superposition of relativistic momenta with respect to the laboratory frame. Here, we will argue that this approach can be extended to superluminal quantum reference frames. 

The spin observables of a particle $A$ with spin $s_A$ are well defined via the Pauli matrices $\hat{\sigma}^i_{s_A}$ in its rest frame. In order to find the spin observable of particle $A$ in the laboratory frame $C$, we boost this observable (which boils down to applying the parity-swap operator as we are changing reference frames from $A$ to $C$)
\begin{equation}
    \hat{P}_{AC} \hat{\sigma}^i_{s_A} \hat{P}_{AC}^\dagger.
\end{equation}
On the other hand, the spin observable for particle $B$ in the rest frame of $A$ is given by
\begin{gather}
\begin{aligned}
    \int &d\bar{p}_B d\bar{p}'_B \eta(p_B) \bar{\eta}(p'_B) \hat{U}(L_{p_B})\hat{\sigma}^i_{s_B}\hat{U}(L_{p'_B})^\dagger \\
    &= \hat{U}_{AB} \hat{\sigma}^i_{s_B}\hat{U}^\dagger_{AB}
\end{aligned}
\end{gather}
where $\eta(p_B)$ is a weight function, encoding the information about the momentum of $B$ as seen from $A$ and we introduced $\hat{U}_{AB} := \int d\bar{p}_B \eta(p_B) \hat{U}(L_{p_B})$. After being boosted to the lab frame the above yields
\begin{gather}
\begin{aligned}
    \hat{U}(L_{\hat{p}_C}) \hat{U}_{AB} \hat{\sigma}^i_{s_B}\hat{U}^\dagger_{AB} \hat{U}(L_{\hat{p}_C})^\dagger = \hat{U}_{BC} \hat{\sigma}^i_{s_B}\hat{U}^\dagger_{BC}
\end{aligned}
\end{gather}
where we introduced $\hat{U}_{BC}:= \hat{U}(L_{\hat{p}_C}) \circ \hat{U}_{AB}$.


Note that the boost to the lab frame may correspond to subluminal, superluminal or even superpositions of both types of velocities. We can combine the above and add measurement settings $\boldsymbol{x}, \boldsymbol{y}$ to obtain the overall Bell measurement in the laboratory frame

\begin{equation}\label{boostobs}
    \hat{P}_{AC}(\boldsymbol{x} \cdot \hat{\boldsymbol{\sigma}}_{s_A} \otimes \boldsymbol{y} \cdot \hat{U}_{BC} (\hat{\boldsymbol{\sigma}}_{s_B} \otimes \mathbb{1}_C)\hat{U}_{BC}^\dagger )\hat{P}_{AC}^\dagger.
\end{equation}

We can similarly obtain the state of a particle from its rest frame description $\ket{(m_A, 0),  z}$ where $z = \pm 1$\footnote{One could choose the convention that each particle considers itself to be in the up state as is done in \cite{de_la_Hamette_2020, cepollaro2024sum}. Here, we do not make this assumption, following instead \cite{streiter2021relativistic}} refers to the spin eigenstates along the $z$-axis.
For the states of the particle, we need to be somewhat more careful and account for the fact that states of the two particles can be entangled. Hence, we need to decompose the overall entangled state using the spin basis states. In the rest frame of $A$, we then have
\begin{gather}
\begin{aligned}
    \sum_{z, z' = \pm 1} \lambda_{zz'} &\ket{(m_A, 0),  z}_{A s_A} \otimes \\&\int d\bar{p}_B \eta(p_B) \hat{U}(L_{p_B}) \ket{(m_B, 0),  z'}_{B s_B} \otimes \ket{\phi}_C.
\end{aligned}
\end{gather}
We can then boost this to the rest frame of $C$
\begin{gather}\label{boostspin}
\begin{aligned}
    \sum_{z, z' = \pm 1} &\lambda_{zz'} \hat{P}_{AC}( \ket{(m_A, 0),  z}_{A s_A} \otimes \\\int& d\bar{p}_C d\bar{p}_B \eta(p_B) \hat{U}(L_{p_C}) \hat{U}(L_{p_B}) \ket{(m_B, 0),  z'}_{B s_B} \otimes \\&\ket{p_C} \braket{p_C|\phi}_C) \\
    =& \sum_{z, z' = \pm 1} \lambda_{zz'} \hat{P}_{AC}( \ket{(m_A, 0),  z}_{A s_A} \otimes \\&\hat{U}_{BC} (\ket{(m_B, 0),  z'}_{B s_B} \otimes \ket{\phi}_C)).
\end{aligned}
\end{gather}
where $\ket{\phi}_C$ is some arbitrary state of the laboratory.


If we now calculate the probability of each measurement outcome between \cref{boostspin,boostobs}, we can use unitarity of the QRF transformations, $\hat{P}_{AC}^\dagger \hat{P}_{AC}$ and $\hat{U}_{BC}^\dagger \hat{U}_{BC}$ which are equal to the identities on the appropriate spaces. We are then left with the probabilities for a Bell test with two particles at rest. Hence, these probabilities are independent of the reference frame, including superluminal reference frames, as the QRF transformations we used were arbitrary.

Notably, however, while the probabilities are invariant, the interpretation of this experiment might not be. The relation between the spatial locations, for example, can be such that Alice is no longer spacelike separated from Bob and/or the setting $x$ ($y$) may now be in the past of the outcome $a$ ($b$).

\section{Discussion}

We have shown how to extend quantum reference frames (QRF) to superluminal Lorentz transformations and as exemplary applications of our conceptual result we have shown how to cast a number of issues with superluminal particles and observers in a consistent manner. A novelty of our extension is that these issues manifest at the level of quantum superposition and, perhaps more interestingly, entanglement as well. This is particularly relevant for the ideas proposed by \cite{dragan2020quantum} that superluminal observers lead to fundamental uncertainty. In such a world, no such thing as a truly classical system exists, classicality being only an approximation\footnote{See the reply to counterargument 2 in \cite{dereply} where the authors explicitly clarify that this is indeed their point of view}; a perspective that also underlies the QRF framework itself \cite{giacomini2019quantum}). Consequently, it is not sufficient to resolve potential problems with superluminal observers only at the level of classical reference frame transformations. 





The energy problem of tachyons has also been tackled in \cite{paczos2024covariant} from a quantum field theoretic perspective. The authors (similar to \cite{schwartz2018tachyon}) extend a Fock space $F$ to the Fock space $F \otimes F^*$, where $F^*$ is the dual of the first space, the interpretation being that subluminal Lorentz transformation do not keep the labels ``incoming'' and ``outgoing'' of tachyons invariant. As we have seen, this problem can equally be tackled in (1+1) dimensions with a quantum reference frame framework extended to the superluminal regime. Unlike \cite{schwartz2018tachyon,paczos2024covariant} our framework extends this solution to cover negative energies arising from superluminal Lorentz transformations and, perhaps more importantly, also to the case where particles may be in superposition of negative and positive energies after a QRF transformation. This framework will be easier to use for studying superluminal Lorentz transformations using  quantum information theoretic tools instead of the quantum field theoretic approach of \cite{schwartz2018tachyon,paczos2024covariant}. 



Spin and momentum couple in relativistic systems, raising questions about Bell violations in the relativistic regime. However, by extending the QRF framework, we demonstrate that Bell values remain invariant even for reference frames with superluminal boosts. 

However, the fact that superluminal boosts can change the ordering between events and the labels ``incoming'' and ``outgoing'' reveals a tension between superluminal observers and operational approaches which could thus have ramifications for the interpretation of a Bell experiment. In the reference frame of a superluminal observer, a setting can end up in the causal past of an outcome it is supposed to influence. If settings are assumed to be free variables (i.e., uncorrelated to anything in their causal past) and we do not allow backwards-in-time signalling, we appear to arrive at a contradiction. Further, does the notion of a free variable make sense if whether a variable is free or not becomes observer-dependent? These interpretational questions are further complicated by the fact that we could transform, for example, into the reference frame of an observer whose velocity is in superposition between subluminal and superluminal speeds. We note that these operational concerns have not been brought up by previous works \cite{schwartz2018tachyon,dragan2020quantum,paczos2024covariant} which were also motivated by the idea of making tachyons and/or superluminal observers consistent with important physical principles like energy positivity, no-backwards-in-time signalling or no superluminal signalling. Analysing this problem in detail will be subject of a future work.


The work of \cite{dragan2020quantum} has also been challenged by a number of other works \cite{horodecki2023comment,jodlowski2024covariant,del2023comment,grudka2022comment}. While our work shows that superluminal observers can be described consistently using quantum reference frames from the point of view of energy, this still leaves the possibility that they run afoul of some other physical principles.

In another follow-up work \cite{sen2026superluminal}, we examine whether the indeterminacy we get from including superluminal transformations is the same indeterminacy that arises from traditional classical theories.\\

The authors of \cite{dragan2020quantum} linked the two pillars of physics—quantum theory and relativity—suggesting that the latter serves as the foundation for the former. Here, we put superluminal Lorentz transformations (aspect of the latter) into the quantum reference frames framework and lay down a framework from a more foundational perspective, rather than the usual field-theoretic approach used in discussion of tachyons. This result highlights a perspective of connecting these theories that has not been explored before.\\ 




\begin{acknowledgements}
     AS and {\L}R acknowledge IRA Programme, project no. FENG.02.01-IP.05 0006/23, financed by the FENG program 2021-2027, Priority FENG.02, Measure FENG.02.01., with the support of the FNP. MS is supported by the National Science Centre, Poland (Opus project, Categorical Foundations of the Non-Classicality of Nature, project no. 2021/41/B/ST2/03149). We are indebted to Pawe{\l} Cie{\'s}li{\'n}ski for valuable comments. AS would like to thank Flavio del Santo for suggesting relevant citations regarding the group structure of superluminal Lorentz transformations. 
\end{acknowledgements} 

\bibliography{bibliography}

\section*{Appendix}

The subluminal Lorentz transformation is given by: 
\[
 \begin{bmatrix} x' \\ t' \end{bmatrix}
 =
   \begin{bmatrix}
   \frac{1}{\sqrt{1-\frac{V^2}{c^2}}} & \frac{-V}{\sqrt{1-\frac{V^2}{c^2}}}\\
   \frac{-V/c^2}{\sqrt{1-\frac{V^2}{c^2}}} & \frac{1}{\sqrt{1-\frac{V^2}{c^2}}} 
   \end{bmatrix}
\begin{bmatrix}x\\t \end{bmatrix}
\]
which can also be written as:
$$L_{sub}=\left(\begin{array}{c|c}
\frac{\hat{p}_0}{m c} & -\frac{\hat{p}_i}{m c} \\
\hline-\frac{\hat{p}_i}{m c} & \delta_{i j}+\frac{\hat{p}_i \hat{p}_j}{m c\left(\hat{p}_0+m c\right)}
\end{array}\right)$$
Whereas the superluminal Lorentz transformation matrix is given by \cite{dragan2020quantum}:
\[
 \begin{bmatrix} x' \\ t' \end{bmatrix}
 =
 \pm \frac{V}{|V|}
  \begin{bmatrix}
   \frac{1}{\sqrt{\frac{V^2}{c^2}-1}} & \frac{-V}{\sqrt{\frac{V^2}{c^2}-1}}\\
   \frac{-V/c^2}{\sqrt{\frac{V^2}{c^2}-1}} & \frac{1}{\sqrt{\frac{V^2}{c^2}-1}} 
   \end{bmatrix}
\begin{bmatrix}x\\t \end{bmatrix}
\] 
which again can be written as: 

$$L_{sup}=\pm \frac{\hat{p}_i}{|\hat{p}_i|}\left(\begin{array}{c|c}
\frac{\hat{p}_0}{m c} & -\frac{\hat{p}_i}{m c} \\
\hline-\frac{\hat{p}_i}{m c} & \delta_{i j}+\frac{\hat{p}_i \hat{p}_j}{m c\left(\hat{p}_0+m c\right)}
\end{array}\right)$$

Now, in the rest frame, the spin observables satisfy the SU(2) algebra for the spin and can be operationally defined by the Stern Gerlach experiment \cite{giacomini2019relativistic}. We now need a definition of spin transformation corresponding to the superposition of Lorentz boosts (including superluminal Lorentz boosts) to the QRF of the laboratory. 

We will now consider three kinds of transformations and comment on the Bell violations in each case: i) subluminal to subluminal reference frame; ii) subluminal to superluminal refernce frame; iii) superluminal to superluminal reference frame. Now, the case iii), as of now has no physical existence if we only consider relational quantities, since there is no way it can be  verified if our reference frame is a superluminal reference frame. Case i) is discussed comprehensively by\cite{ballesteros2021group}. The case ii) is the subject of concern for this paper. 

 Let us try to construct the Pauli-Lubanski spin operator $\Sigma_p$ for $\tilde{L}_p$ where $\tilde{L}_p$ includes Lorentz transformations for both subluminal and superluminal speeds. For this we follow \cite{giacomini2019relativistic}and define a generic basis element $\hat{U}(\tilde{L}_p) |k, \vec{\sigma}\rangle= |p, \Sigma_p\rangle$ where $|k, \vec{\sigma} \rangle = \Sigma_{\lambda}c_{\lambda} |k, \lambda \rangle$ is represented in spin basis and $\hat{U}(\tilde{L}_p)$ is any Lorentz boost from rest frame to the frame with momentum p. Note that the transformation $\hat{U}(\tilde{L}_p)$ will have a parity part and an unitary part $S_{L}$. To describe the behaviour of the transformations $S_{L}$ dictates the structure of the unitary while the parity operator dictates the direction. 

The Pauli-Lubanski operator acts in the following way:

\[\begin{aligned}
\hat{\Sigma}_{\hat{p}}^\mu\left|p, \Sigma_p\right\rangle
& =\hat{\Sigma}_p^\mu \hat{U}\left(\tilde{L}_p\right)|k, \vec{\sigma}\rangle\\
& =\hat{U}\left(\tilde{L}_p\right) \hat{U}^{\dagger}\left(\tilde{L}_p\right) \hat{\Sigma}_p^\mu \hat{U}\left(\tilde{L}_p\right)|k, \vec{\sigma}\rangle \\
& =\hat{U}\left(\tilde{L}_p\right)\left(\tilde{L}_{-p}\right)_\nu^\mu \hat{\Sigma}_p^\nu|k, \vec{\sigma}\rangle\\
&=\sum_\lambda c_\lambda \hat{U}\left(L_p\right)\left(L_{-p}\right)_\nu^\mu \hat{\sigma}^\nu|k, \lambda\rangle\\
& =\sum_{\lambda, \lambda^{\prime}} c_\lambda \hat{U}\left(\tilde{L}_p\right)\left(\tilde{L}_{-p}\right)_\nu^\mu\left[\sigma^\nu\right]_{\lambda^{\prime} \lambda}\left|k, \lambda^{\prime}\right\rangle \end{aligned}\]\\ 

Now, we need to find out the relationship between $\hat{U}\left(\tilde{L}_p\right)$ and $\hat{U}\left(\tilde{L}_{-p}\right)$. We noted before that $L_{sup}(-v)$ and $-L_{sup}(v)$ is unitary conjugate but not an even function like the subluminal lorentz transformation that gives in to this confusion. If we look into the structure of unitaries, the structure of the unitaries $\hat{U}\left(\tilde{L}_p\right)$ and $S_L$ dictates us that the only possible relation with the direction of momentum is $\hat{U}\left(\tilde{L}_-p\right) =\hat{U}^{\dagger}\left(\tilde{L}_p\right)$, hence resuming the calculations to be:
\begin{align*}
&\sum_{\lambda, \lambda^{\prime}} c_\lambda \hat{U}(\tilde{L}_p) \left( \tilde{L}_{-p} \right)_\nu^\mu \left[ \sigma^\nu \right]_{\lambda^{\prime} \lambda} \left| k, \lambda^{\prime} \right\rangle \\
&= \sum_{\lambda, \lambda^{\prime}} c_\lambda \left( \tilde{L}_{-p} \right)_\nu^\mu \left[ \sigma^\nu \right]_{\lambda^{\prime} \lambda} \left| p, \Sigma_p(\lambda^{\prime}) \right\rangle \\
&= \left( \tilde{L}_{-p} \right)_\nu^\mu \hat{\sigma}^\nu \left| p, \Sigma_p \right\rangle .
\end{align*}

This looks exactly the same as the subluminal case, and hence as per \cite{streiter2021relativistic}, we get frame independent Bell inequalities even for superluminal and subluminal-superluminal mixed cases.


\end{document}